# Bipartite quantum channels using multipartite cluster-type entangled coherent states

P. P. Munhoz,[1] J. A. Roversi,[2] A. Vidiella-Barranco,[2] and F. L. Semião[1]

[1]*Departamento de Física, Universidade Estadual de Ponta Grossa,
Campus Uvaranas, 84030-900 Ponta Grossa, PR, Brasil*

[2]*Instituto de Física "Gleb Wataghin", Universidade Estadual de Campinas,
Unicamp, 13083-970 Campinas, SP, Brasil*

We propose a particular encoding for bipartite entangled states derived from multipartite cluster-type entangled coherent states (CTECSs). We investigate the effects of amplitude damping on the entanglement content of this bipartite state, as well as its usefulness as a quantum channel for teleportation. We find interesting relationships among the amplitude of the coherent states constituting the CTECSs, the number of subsystems forming the logical qubits (redundancy), and the extent to which amplitude damping affects the entanglement of the channel. For instance, in the sense of sudden death of entanglement, given a fixed value of the initial coherent state amplitude, the entanglement life span is shortened if redundancy is increased.

## I. INTRODUCTION

In an early stage of quantum mechanics, entanglement was mainly related to fundamental questions [1]. More recently, due to seminal articles on dense coding [2] and teleportation [3], it began to be recognized as a resource for performing communication tasks. Since then, entanglement and its properties have been deeply investigated and several applications in many contexts have been found. Most of the knowledge built so far concerns the bipartite scenario, but it is well known that a full understanding of multipartite entanglement is required if we want to make the most of quantum correlations. One of the main differences between bipartite and multipartite entanglement is the existence of inequivalent classes of entanglement [4]. One-way quantum computing is a particular instance of application of multipartite entanglement [5]. This computation model is based on local measurements on an initially prepared highly entangled multipartite state called a cluster state [6]. Several schemes for cluster state generation have then been suggested, using different physical settings. We may cite proposals in linear optics and spontaneous down-conversion [7], cavity QED [8], hybrid cavity QED with linear optics [9], trapped ions [10], and superconducting qubits [11], just to name a few.

A natural development in the study of entangled states was the introduction of nonorthogonal states for the subsystems, in special bosonic coherent states. It is important to remark that such entangled states had previously appeared in the rich quantum optics literature [12]. After the work of Sanders [13], such states began to be referred to as entangled coherent states (ECSs). In [14], the entanglement properties of the Bell-type ECSs, also called the quasi-Bell states, were discussed. A remarkable fact pointed out in [14] is that some of those quasi-Bell states are in fact maximally entangled in $\mathbb{C}^2 \otimes \mathbb{C}^2$, irrespective of the amplitude of the coherent states. More recently, similar properties belonging to multipartite ECSs have been discussed [15]. It is also important to mention that encoding in finite-dimensional spaces in terms of coherent states has also been previously considered for teleportation [16, 17], Bell inequalities violation [18], and entanglement purification [19].

Different constructions of cluster-type ECSs were independently proposed in [20] and [21]. The construction presented in [21] has motivated the appearance of many generation schemes in the literature, specially in cavity QED [22] and the traveling optical fields domain [23]. Quite recently, applications of cluster-type entangled coherent states (CTECSs) for quantum communication have also appeared [24, 25]. Given this strong interest in the CTECSs, we consider here a special codification procedure where a CTECS involving $2p$ subsystems is regarded as an encoded bipartite entangled; that is, it is shared by two parties. We then analyze the bipartite entanglement shared between the two parties, as well as the reliability for teleportation of the established quantum channel.

The article is organized as follows. In Sec. II we describe a particular encoding using CTECSs which leads to the establishment of a quantum channel shared by two parties. In Sec. III we consider amplitude damping in this channel. In Sect. IV we follow the approach presented in [16] to define an orthonormal basis useful for entanglement analysis. In Sec. 5 we analyze the bipartite entanglement and also the fidelity of teleportation. In Sec. 6 we summarize our results and conclude.

## II. LOGICAL QUBITS ENCODING

We consider now a special instance of multimode ECSs, which were introduced as CTECSs in [21]. Such states may be considered as belonging to the $4^p$-dimensional space state of qubits $(\mathbb{C}^2)^{\otimes 2p}$, and they may be written as

$$|\text{CTECS}_{2p}\rangle = N^{1/2}(|\beta\rangle^{\otimes p}|\beta\rangle^{\otimes p} - z_p|\beta\rangle^{\otimes p}|-\beta\rangle^{\otimes p} \\ - z_p|-\beta\rangle^{\otimes p}|\beta\rangle^{\otimes p} - z_p^2|-\beta\rangle^{\otimes p}|-\beta\rangle^{\otimes p}), \tag{1}$$

where $\beta$ is the complex amplitude of a coherent state ($\beta$ is an eigenvalue of the bosonic annihilation operator), $N = \frac{1}{2}[2 + (1-z_p^2)c^2]^{-1}$ is the normalization constant, $z_p = (-i)^p$ is a relative phase, and $c$ is the overlap,

$$\langle \beta_p | -\beta_p \rangle = e^{-2|\beta|^2 p} \equiv c, \tag{2}$$

between $|\pm\beta_p\rangle \equiv |\beta\rangle^{\otimes p} = |\pm\beta, \ldots, \pm\beta\rangle \equiv |\pm\beta\rangle_1 \otimes \ldots \otimes |\pm\beta\rangle_p$. With this definition, we have introduced a quasiorthog-

onal encoded logical basis which allows us to rewrite (1) as

$$|\text{CTECS}_{2p}\rangle = N^{1/2}(|\beta_p, \beta_p\rangle - z_p|\beta_p, -\beta_p\rangle \\ - z_p|-\beta_p, \beta_p\rangle - z_p^2|-\beta_p, -\beta_p\rangle), \quad (3)$$

which we will assume to be a bipartite entangled state share by two parties.

Physically speaking, each party possesses $p$ subsystems, and these $p$ pairs of subsystems are collectively described by the CTECSs. In the next sections, we will analyze the usefulness of the bipartite state (3) as a quantum channel shared by two parties, under a more realistic situation where damping had turned it into a mixed state. As a last remark about (3), it is easy to see that for $p$ even the normalization constant does not depend on the coherent amplitude $\beta$. However, the amplitude $\beta$ is relevant in another respect. Depending on the choices for the coherent amplitude $\beta$ and the number of pairs $p$ in the encoding, the logical ket basis may be effectively considered as mutually orthogonal for any practical purposes. For instance, if $|\beta| = 3$ and $p = 1$ or if $|\beta| = 1$ and $p = 9$, the overlap is just about $10^{-8}$. Such an interplay between $p$ and $\beta$, which are completely independent quantities, leads to interesting results when damping is included. This is the subject we are now going to treat.

### III. AMPLITUDE DAMPING

In any real situation one might expect to find complicated decoherence mechanisms which are in general not easy to model. Two important decoherence mechanisms are, for instance, the amplitude and phase damping [26], usually treated within the master equation framework [27]. As a first account of decoherence, we will consider here the effects of the amplitude-damping mechanism acting on the quantum channel (3). Each subsystem will suffer the action of the coupling to a vacuum environment. This may be modeled via a beam-splitter transformation [28],

$$|\beta\rangle_S |0\rangle_E \to |\beta\sqrt{\eta}\rangle_S |\beta\sqrt{1-\eta}\rangle_E, \quad (4)$$

where $S$ and $E$ stand for system and environment, respectively, $\eta$ (transmissivity) is a real parameter ranging from zero to one, and the kets refer to coherent or vacuum states. An example of a physical situation which may be modeled by a beam-splitter-type interaction is an optical field mode crossing an optical fiber. The beam-splitter transmissivity in this case will be an exponential energy loss during transmission through the fiber, i.e., $\eta = e^{-\lambda L}$, where $\lambda$ is the fiber loss coefficient and $L$ the transmission distance [28]. From this example we then note that $\eta$ can be regarded as a physical parameter related to the amplitude-damping process. Another example is given by an electromagnetic cavity field mode initially prepared in a coherent state. For non ideal cavities, photon leakage through the walls follows a time evolution given by $\eta = e^{-\kappa t}$, where $\kappa$ represents the photon leakage rate.

Following the general description (4), we now study the effect of damping, characterized by the exponential function $\tau \equiv \eta = e^{-\kappa t}$, on the initial state (3). Due to the interaction with the environment, the initially pure quantum channel (3) evolves to a mixed state. Turning back to the beam-splitter transformation given previously (4), we trace out the environment and obtain its action on each partition of (3):

$$|\beta_p\rangle\langle\beta_p'| \to \tilde{c}|\tilde{\beta}_p\rangle\langle\tilde{\beta}_p'|, \quad (5)$$

where $\tilde{c} = c^{r^2}$, $|\tilde{\beta}_p\rangle = |\beta_p\sqrt{\tau}\rangle$, $\tau = e^{-\kappa t}$ and $r$ is a normalized parametrization of time, related to $\tau$ as $r = \sqrt{1-\tau}$ [16]. Hence, it follows that for $t = 0$ we have $\tau = 1$ and $r = 0$, while for $t \to \infty$, $\tau \to 0$ and $r \to 1$. It is important to note that the same result would have been obtained from the usual master equation approach [16, 18].

Therefore, the ideal quantum channel (3) under the action of the amplitude-damping mechanism evolves to the following mixed ECS:

$$\varrho(t) = N(|\tilde{\beta}_p, \tilde{\beta}_p\rangle\langle\tilde{\beta}_p, \tilde{\beta}_p| - z_p^*\tilde{c}|\tilde{\beta}_p, \tilde{\beta}_p\rangle\langle\tilde{\beta}_p, -\tilde{\beta}_p| \\ - z_p^*\tilde{c}|\tilde{\beta}_p, \tilde{\beta}_p\rangle\langle-\tilde{\beta}_p, \tilde{\beta}_p| - z_p^2\tilde{c}^2|\tilde{\beta}_p, \tilde{\beta}_p\rangle\langle-\tilde{\beta}_p, -\tilde{\beta}_p| \\ - z_p\tilde{c}|\tilde{\beta}_p, -\tilde{\beta}_p\rangle\langle\tilde{\beta}_p, \tilde{\beta}_p| + |\tilde{\beta}_p, -\tilde{\beta}_p\rangle\langle\tilde{\beta}_p, -\tilde{\beta}_p| \\ + |z_p|^2\tilde{c}^2|\tilde{\beta}_p, -\tilde{\beta}_p\rangle\langle-\tilde{\beta}_p, \tilde{\beta}_p| + z_p^*\tilde{c}|\tilde{\beta}_p, -\tilde{\beta}_p\rangle \\ \times\langle-\tilde{\beta}_p, -\tilde{\beta}_p| - z_p\tilde{c}|-\tilde{\beta}_p, \tilde{\beta}_p\rangle\langle\tilde{\beta}_p, \tilde{\beta}_p| \\ + |z_p|^2\tilde{c}^2|-\tilde{\beta}_p, \tilde{\beta}_p\rangle\langle\tilde{\beta}_p, -\tilde{\beta}_p| + |-\tilde{\beta}_p, \tilde{\beta}_p\rangle\langle-\tilde{\beta}_p, \tilde{\beta}_p| \\ + z_p^*\tilde{c}|-\tilde{\beta}_p, \tilde{\beta}_p\rangle\langle-\tilde{\beta}_p, -\tilde{\beta}_p| - z_p^2\tilde{c}^2|-\tilde{\beta}_p, -\tilde{\beta}_p\rangle \\ \times\langle\tilde{\beta}_p, \tilde{\beta}_p| + z_p\tilde{c}|-\tilde{\beta}_p, -\tilde{\beta}_p\rangle\langle\tilde{\beta}_p, -\tilde{\beta}_p| + z_p\tilde{c} \\ \times|-\tilde{\beta}_p, -\tilde{\beta}_p\rangle\langle-\tilde{\beta}_p, \tilde{\beta}_p| + |-\tilde{\beta}_p, -\tilde{\beta}_p\rangle\langle-\tilde{\beta}_p, -\tilde{\beta}_p|). \quad (6)$$

It is now easy to see that in the limit of $r \to 1$ (infinite time), state (6) becomes $|0\rangle^{\otimes p}|0\rangle^{\otimes p}$, completely losing its entanglement. What is more interesting, though, is the fact that depending on the values of the amplitude $\beta$ and the number of redundant physical qubits $p$, disentangling may take place at finite times long before the channel is transformed into $|0\rangle^{\otimes p}|0\rangle^{\otimes p}$. This is an example of the well-discussed phenomenon of entanglement sudden death [29]. These results will be shown in the next sections.

### IV. ORTHONORMAL BASIS

The coherent state basis considered up to now is, strictly speaking, formed by nonorthogonal states [see Eq. (2)]. For evaluation of entanglement and fidelity of teleportation, it is useful to span the system density operator on an orthogonal basis. It is important to remark that there is no preferred orthogonal basis to choose, because they all lead to the same result [15]. In this work, we use the even and odd coherent states [30] given by

$$|\beta^{\pm}\rangle = M_{\pm}^{1/2}(|\beta\rangle \pm |-\beta\rangle), \quad (7)$$

where $M_{\pm} = \frac{1}{2}(1 \pm e^{-2|\beta|^2})^{-1}$ are normalization constants.

For our purposes, we need the multimode generalization of the even and odd coherent states [12],

$$|\beta_p^{\pm}\rangle \equiv M_{\pm,p}^{1/2}(|\beta_p\rangle \pm |-\beta_p\rangle), \quad (8)$$

where $M_{\pm,p} = \frac{1}{2}(1 \pm c)^{-1}$ are the normalization constants. Please notice that the notation may induce a misinterpretation, that is, $|\beta_p^\pm\rangle \neq |\beta^\pm\rangle^{\otimes p}$. If the right definition (8) is kept in mind, there will be no trouble hereafter. We can now define the orthonormal kets,

$$|\tilde{\beta}_p^\pm\rangle \equiv \tilde{M}_{\pm,p}^{1/2}(|\tilde{\beta}_p\rangle \pm |-\tilde{\beta}_p\rangle), \qquad (9)$$

where $\tilde{M}_{\pm,p} = \frac{1}{2}(1 \pm c^{1-r^2})^{-1}$. Although the basis is now time-dependent, orthogonality is maintained at all times, even for $t \to \infty$ [18]. It is straightforward to show that the logical qubits $|\pm\tilde{\beta}_p\rangle$ can be written in terms of $\{|\tilde{\beta}_p^+\rangle, |\tilde{\beta}_p^-\rangle\}$ as

$$|\pm\tilde{\beta}_p\rangle = \tilde{a}|\tilde{\beta}_p^+\rangle \pm \tilde{b}|\tilde{\beta}_p^-\rangle, \qquad (10)$$

where $\tilde{a} = \frac{1}{2}\tilde{M}_{+,p}^{-1/2}$, and $\tilde{b} = \frac{1}{2}\tilde{M}_{-,p}^{-1/2}$, with $|\tilde{a}|^2 + |\tilde{b}|^2 = 1$.

We are now finally in position to obtain the matrix representing the quantum channel (6) in the orthonormal basis $\{|\tilde{\beta}_p^+, \tilde{\beta}_p^+\rangle, |\tilde{\beta}_p^+, \tilde{\beta}_p^-\rangle, |\tilde{\beta}_p^-, \tilde{\beta}_p^+\rangle, |\tilde{\beta}_p^-, \tilde{\beta}_p^-\rangle\}$. One can show that it reads

$$\varrho(t) = \begin{pmatrix} \tilde{a}^4 & -i^p\tilde{a}^3\tilde{b}\tilde{c} & -i^p\tilde{a}^3\tilde{b}\tilde{c} & -\tilde{a}^2\tilde{b}^2\tilde{c}^2 \\ -i^p\tilde{a}^3\tilde{b}\tilde{c} & \tilde{a}^2\tilde{b}^2 & \tilde{a}^2\tilde{b}^2\tilde{c}^2 & i^p\tilde{a}\tilde{b}^3\tilde{c} \\ -i^p\tilde{a}^3\tilde{b}\tilde{c} & \tilde{a}^2\tilde{b}^2\tilde{c}^2 & \tilde{a}^2\tilde{b}^2 & i^p\tilde{a}\tilde{b}^3\tilde{c} \\ -\tilde{a}^2\tilde{b}^2\tilde{c}^2 & i^p\tilde{a}\tilde{b}^3\tilde{c} & i^p\tilde{a}\tilde{b}^3\tilde{c} & \tilde{b}^4 \end{pmatrix} \quad (11)$$

for $p$ even and

$$\varrho(t) = \frac{\tilde{a}^2\tilde{b}^2}{1+c^2}$$
$$\times \begin{pmatrix} \frac{\tilde{a}^2}{\tilde{b}^2}(1+\tilde{c}^2) & 0 & 0 & 2i^p\tilde{c} \\ 0 & (1-\tilde{c}^2) & 0 & 0 \\ 0 & 0 & (1-\tilde{c}^2) & 0 \\ -2i^p\tilde{c} & 0 & 0 & \frac{\tilde{b}^2}{\tilde{a}^2}(1+\tilde{c}^2) \end{pmatrix}$$
$$(12)$$

for $p$ odd. Now, it is interesting to notice that the case $p$ odd is special because it is an instance of $X$ states. These states have some peculiar properties that are discussed in [31].

With these matrices, we now proceed to analyze the bipartite entanglement content and the extent to which the quantum channel may be considered for teleportation.

## V. ENTANGLEMENT STUDY

A convenient way to study bipartite entanglement is through the concurrence. Once we have rewritten the state (6) in the form (11) and (12), we may calculate the concurrence using standard procedures. For even $p$, the matrix has in general no null elements and then we must follow the general recipe [32]

$$C = \max(0, \lambda_1 - \lambda_2 - \lambda_3 - \lambda_4), \qquad (13)$$

where the parameters $\lambda_i$ ($i = 1, \ldots, 4$) are the square roots of the eigenvalues (in decreasing order) of the non-Hermitian operator $\varrho\tilde{\varrho}$, written in the same basis (11), and

$$\tilde{\varrho} = (\sigma^y \otimes \sigma^y)\varrho^*(\sigma^y \otimes \sigma^y) \qquad (14)$$

is the spin-flipped operator, $\varrho^*$ being the complex conjugate of (11). Performing the preceding calculations, the concurrence for $p$ even will read

$$C = 2\tilde{a}^2\tilde{b}^2 \max[0, \tilde{c}^2 + 2\tilde{c} - 1]. \qquad (15)$$

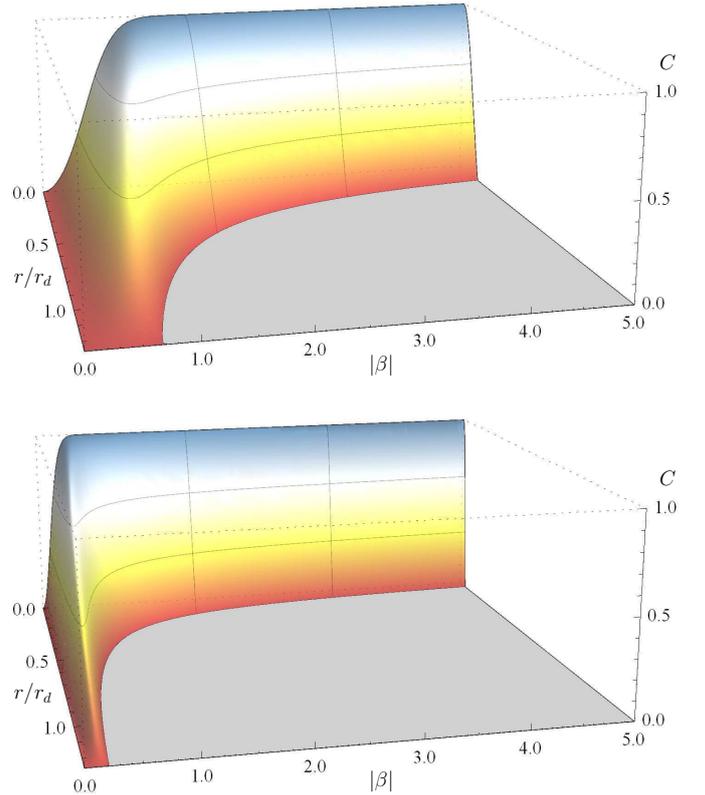

FIG. 1. (Color online) Concurrence as a function of the renormalized time parameter $r/r_d$ and the initial coherent amplitude $\beta$ for $p = 1$ (top) and $p = 10$ (bottom).

For odd $p$, the related matrix is much simpler (many zeros) and we may easily obtain the concurrence from general results found in [31]:

$$C = \frac{2\tilde{a}^2\tilde{b}^2}{1+c^2} \max[0, \tilde{c}^2 + 2\tilde{c} - 1]. \qquad (16)$$

We now plot the concurrence as a function of the initial coherent amplitude $\beta$ and the renormalized time parameter $r/r_d$, where $r_d = \sqrt{1 - e^{-1}}$ is the normalized relaxation time parameter. Please notice that this renormalized variable will now take values in the interval $[0, 1/r_d]$. The concurrence plots shown in Fig. 1 reveal the presence of the interesting phenomenon entanglement sudden death (ESD) [29]. From these plots, one can see that ESD happens for coherent amplitudes greater than $\beta$ about 0.7 (0.2), if the number of physical qubits in the encoding of a logical qubit is $p = 1$ ($p = 10$). On the contrary, for sufficiently small values of $\beta$, implying nonorthogonality between $|\beta\rangle$ and $|-\beta\rangle$, complete disentanglement takes place only for $r \to 1/r_d$ (infinite times). In fact, it goes to a multimode vacuum state as already discussed. On the other hand, the greater the initial amplitude $\beta$ the sooner the ESD. By comparing both plots in Fig. 1, one can also see the interesting fact that for weak coherent state amplitudes (close to vacuum), the increase of the redundancy $p$ may be used to increase entanglement in the channel. Such compromise between redundancy and amplitude of the coherent states will become even more evident now that we discuss the fidelity of teleportation.

The maximal fidelity of teleportation which may be obtained

by employing an usual bipartite state as a quantum channel is given by [33]

$$F_{\max} = \frac{2f_{\max} + 1}{3}, \quad (17)$$

where $f_{\max}$ is the fully entangled fraction [34]

$$f_{\max} = \max_{|\psi\rangle}\langle\psi|\varrho|\psi\rangle, \quad (18)$$

with the maximum taken over all bipartite maximally entangled states. Here we follow the same procedure described in [34] to calculate $f_{\max}$. We write the quantum channel (6) in the so-called magic basis $|m_i\rangle$, which in our case is a (time-dependent) encoded basis constituted by the multimode even and odd coherent states:

$$|m_1\rangle = |\Phi^+_{\beta,p}\rangle = \frac{1}{\sqrt{2}}(|\tilde{\beta}^+_p, \tilde{\beta}^+_p\rangle + |\tilde{\beta}^-_p, \tilde{\beta}^-_p\rangle),$$

$$|m_2\rangle = i|\Phi^-_{\beta,p}\rangle = \frac{i}{\sqrt{2}}(|\tilde{\beta}^+_p, \tilde{\beta}^+_p\rangle - |\tilde{\beta}^-_p, \tilde{\beta}^-_p\rangle),$$

$$|m_3\rangle = i|\Psi^+_{\beta,p}\rangle = \frac{i}{\sqrt{2}}(|\tilde{\beta}^+_p, \tilde{\beta}^-_p\rangle + |\tilde{\beta}^-_p, \tilde{\beta}^+_p\rangle),$$

$$|m_4\rangle = |\Psi^-_{\beta,p}\rangle = \frac{1}{\sqrt{2}}(|\tilde{\beta}^+_p, \tilde{\beta}^-_p\rangle - |\tilde{\beta}^-_p, \tilde{\beta}^+_p\rangle). \quad (19)$$

Thus, $f_{\max}$ is simply the highest eigenvalue of the real part of the quantum channel state, when it is written in the encoded magic basis, and reads

$$f_{\max} = \frac{1}{4}[1 + 4\tilde{a}^2\tilde{b}^2\tilde{c}^2 + \sqrt{(\tilde{a}^2 - \tilde{b}^2)^4 + 16\tilde{a}^2\tilde{b}^2\tilde{c}^2}], \quad (20)$$

when $p$ is even, and

$$f_{\max} = \frac{1}{2(1+c^2)}[1 - 2\tilde{a}^2\tilde{b}^2(\tilde{c} - 1)^2 + \tilde{c}^2], \quad (21)$$

when $p$ is odd.

We may now analyze the maximal fidelity of teleportation attainable with the quantum channel (6) as a function of $\beta$ and $r/r_d$. In Fig. 2, we observe that for sufficiently small values of $\beta$, the quantum channel stays useful for teleportation for all times under the action of damping ($F_{\max} > 2/3 \, \forall \, r/r_d$).

It might be interesting to analyze more closely the role played by the redundancy $p$ when the amplitude $\beta$ is fixed. In Fig. 3, we show the maximal fidelity of teleportation for two specific values of $\beta$. It is remarkable that depending on $r/r_d$, which essentially measures the duration of the action of damping on the channel, it is more advantageous to have a large or a small number of subsystems in the encoding (redundancy). This number clearly depends on $\beta$ and $r/r_d$. For example, let us consider the first plot (top) in Fig 3, where we have considered $\beta = 0.5$. One can see that for small values of $r/r_d$, it is more advantageous to have an encoding with high redundancy $p$, while for high degradation of the channel (big values of $r/r_d$), a small redundancy is more appropriate. However, it should be stressed that $r/r_d$ cannot be made arbitrarily small in our analysis due to the fact that the treatment of dissipation is performed in the Markovian approximation.

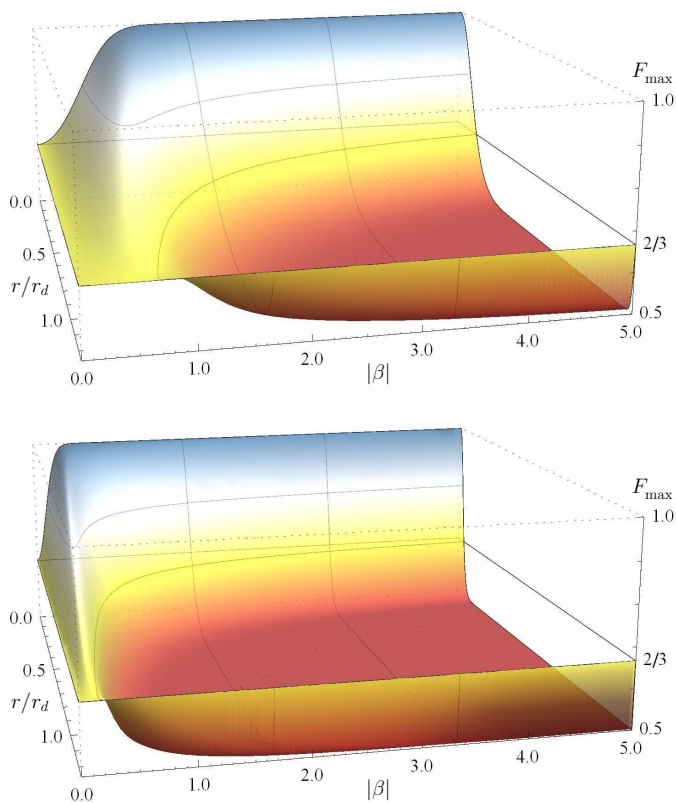

FIG. 2. (Color online) Maximal fidelity of teleportation as a function of the renormalized time parameter $r/r_d$ and the initial coherent amplitude $\beta$ for $p = 1$ (top) and $p = 10$ (bottom).

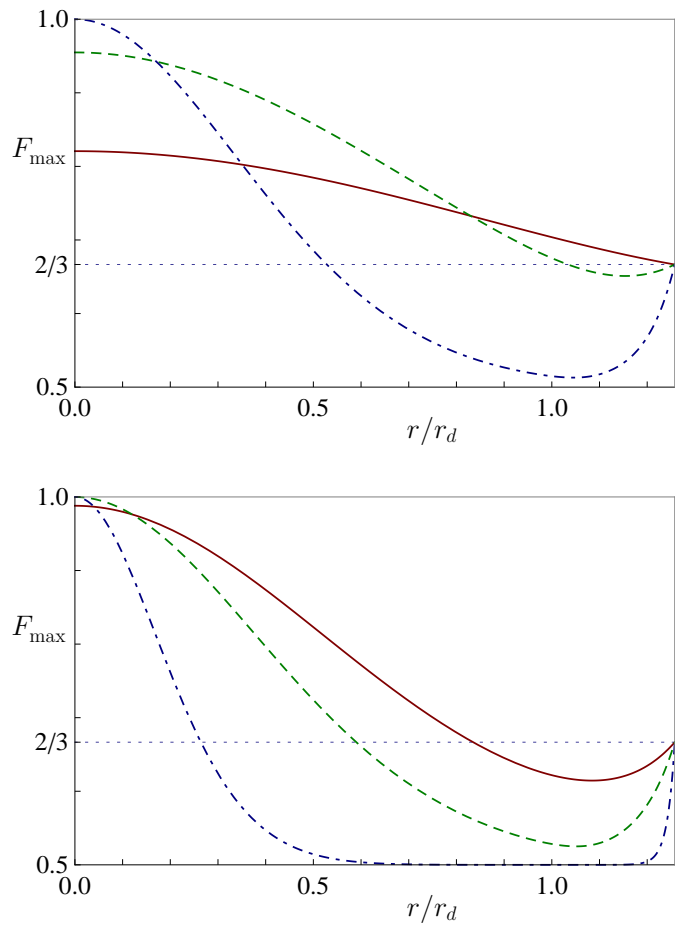

FIG. 3. (Color online) Maximal fidelity of teleportation as a function of the renormalized time parameter $r/r_d$ for initial coherent amplitudes $\beta = 0.5$ (top) and $\beta = 1.0$ (bottom), with $p = 1$ (red solid line), $p = 2$ (green dashed line) and $p = 10$ (blue dot-dashed line).

It is worth noticing that, for odd $p$, one may prove using (16), (17) and (21) that $C = \max[0, 2f_{\max}-1] = \max[0, 3F_{\max}-2]$. In this case, the fidelity of teleportation can be seen as a kind of entanglement detector in the sense that any entangled state leads to a fidelity that cannot be achieved classically ($F_{\max} > 2/3$). Interestingly enough, this result is also valid for the usual Werner state of two qubits [35]. In spite of this, the fidelity of teleportation no longer works as an entanglement detector for even $p$. For example, by considering $p = 2$, $\beta = 0.5$, and $r/r_d = 1.1$, one obtains $C \approx 0.028 > 0$ (entangled state) and $F_{\max} \approx 0.65 < 2/3$ (useless for quantum teleportation). Therefore, there are now situations where the channel is an entangled state, but it yields a fidelity of teleportation lower than the best classical strategy. Such a situation also appears in other quantum systems, for example, in the evolved quantum state of two dipole-dipole coupled qubits under the action of spontaneous emission [36].

## VI. CONCLUSION

We have studied the use of a particular bipartition of a CTECS as an entangled quantum channel after action of amplitude damping. We have constructed an orthonormal basis with multimode even and odd coherent states which allowed us to analyze the entanglement of the encoded state. We have also verified that for ranges of values of the amplitude $\beta$ and the redundancy $p$, entanglement goes abruptly to zero indicating the occurrence ESD. Moreover, in order to find the extent to which such a quantum channel is reliable for quantum information tasks, we have analyzed the maximum fidelity of teleportation. We have also found that the coherent state amplitude and the redundancy in the logical encoding (controlled parameters) may be suitably chosen to increase the fidelity of teleportation of that quantum channel.

## ACKNOWLEDGMENTS

P.P.M. is grateful for the financial support from CNPq (Conselho Nacional de Desenvolvimento Científico e Tecnológico), under Grant No. 141434/2002-3 and from Coordenação de Aperfeiçoamento de Pessoal de Nível Superior (CAPES) and CNPq, under Grant No. PNPD0030082 (Programa Nacional de Pós-Doutorado). J.A.R., A.V.B., and F.L.S. acknowledge the partial support of CNPq (Brazil). This work was performed as part of the Brazilian National Institute of Science and Technology of Quantum Information (INCT-IQ), and CePOF (Optics and Photonics Research Center) FAPESP.